\documentstyle[prl,aps,epsf,floats]{revtex}

\begin{document}

\draft

\twocolumn[\hsize\textwidth\columnwidth\hsize\csname@twocolumnfalse%
\endcsname

\title{$C$-axis Josephson Tunneling Between
YBa$_2$Cu$_3$O$_{7-\delta}$ and Pb: Direct Evidence for Mixed Order
Parameter Symmetry in a High-T$_c$ Superconductor}

\author{K.~A.~Kouznetsov$^1$, A.~G.~Sun$^2$, B.~Chen$^1$,
A.~S.~Katz$^2$, S.~R.~Bahcall$^1$, John Clarke$^1$, R.~C.~Dynes$^2$,
D.~A.~Gajewski$^2$, S.~H.~Han$^2$, M.~B.~Maple$^2$,
J.~Giapintzakis$^3$, J.-T.~Kim$^3$, D.~M.~Ginsberg$^3$.}

\address{{}$^1$Department of Physics, University of California \\
and Materials Sciences Division, Lawrence Berkeley National Laboratory,
Berkeley CA 94720\\
{}$^2$Department of Physics, University of California, 
San Diego, La Jolla, CA 92093\\
{}$^3$Department of Physics,
 University of Illinois at Urbana-Champaign, 
Urbana, IL 61801}

\maketitle

\begin{abstract}
We report a new class of $c$-axis Josephson tunneling experiments in
which a conventional superconductor (Pb) is deposited across a single
twin boundary of a YBa$_2$Cu$_3$O$_{7-\delta}$ crystal.  We measure
the critical current as a function of magnitude and angle of magnetic
field applied in the plane of the junction.  In all samples, we
observe a clear experimental signature of an order parameter phase
shift across the twin boundary.  These results provide strong evidence
for mixed $d$- and $s$-wave pairing in YBCO, with a reversal in the
sign of the $s$-wave component across the twin boundary.


\end{abstract}


]
\def\bottomfraction{.9}
\def\textfraction{.1}

The symmetry of the order parameter in high temperature
superconductors has been a subject of theoretical and experimental
debate since their original discovery \cite{scalapino,reviews}.
Phase-sensitive measurements on YBCO involving currents flowing within
the CuO$_2$ planes, such as corner-junction SQUID experiments
\cite{wollman,brawner,mathai}, corner-junction flux modulation
experiments \cite{wollmanfm,miller}, and tricrystal ring experiments
\cite{tsuei}, have indicated an order parameter $\Delta({\bf k})$ with
primarily $d_{x^2-y^2}$ symmetry under rotations in the plane
[$\Delta({\bf k})\sim \cos k_x^{} a-\cos k_y^{} a$].  The observation,
however, of Josephson tunneling perpendicular to the CuO$_2$ planes
between heavily twinned YBCO and a conventional $s$-wave
superconductor \cite{dynesa,dynesb,losueur,dynesd,kleiner}
demonstrated a significant $s$-wave component.  In this Letter, we
report results from a new $c$-axis Josephson tunneling experiment
which resolves the apparent conflict between the two groups of
experiments and makes a compelling case for mixed order parameter
symmetry in YBCO, with a dominant $d$-wave component and a significant
$s$-wave component.

In the experiment, shown conceptually in Fig.\ 1(a), a $c$-axis
Josephson tunnel junction straddles a single YBCO
twin boundary.  Because Pb is an $s$-wave superconductor, the Pb
counterelectrode couples only to the $s$-wave component of the YBCO
order parameter; the net phase difference with any $d$-wave component
integrates to zero \cite{dynesa}.  If YBCO were a conventional $s$-wave
superconductor, the critical current $I_c({\bf B})$ would exhibit a
Fraunhofer-like dependence on magnetic field strength ${\bf B}$,
applied in the plane of the junction, regardless of the angle of ${\bf
B}$.  If, on the other hand, YBCO were predominantly $d$-wave, any
$s$-wave component to the order parameter would change sign across the
twin boundary \cite{signch}, yielding a dramatically different angular
dependence for $I_c({\bf B})$.  For magnetic fields {\it
perpendicular} to the boundary, $I_c({\bf B})$ would be the same as
for a conventional junction, with a maximum at $B=0$.  Magnetic fields
{\it parallel} to the boundary, however, should produce a local
{minimum} in $I_c$ at $B=0$, as discussed below; this configuration is
analogous to the $d$-wave corner junction experiment
\cite{reviews,wollmanfm}.

\begin{figure}[b]
\epsfysize=8cm\epsfxsize=7cm\centerline{\epsfbox{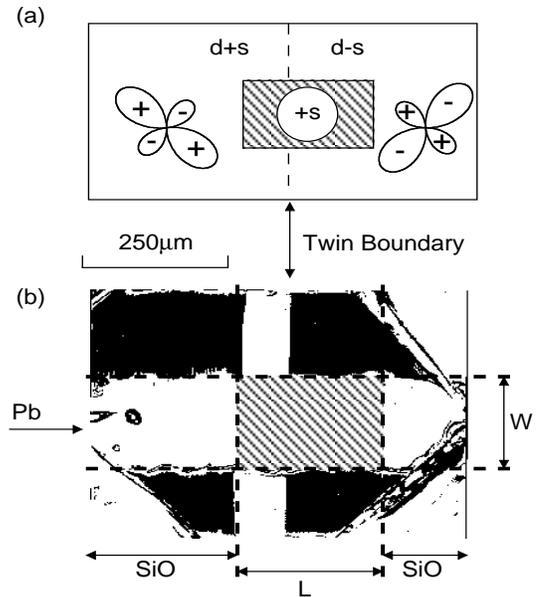}}
 \caption{(a) Schematic diagram of a Pb/YBCO junction (hatched region) grown
across a twin boundary (dashed line).  For the YBCO order parameter
shown, the $s$-wave component changes sign across the boundary.
(b)  Nemarsky microscope photograph of junction B1.
The twin boundary runs vertically through the center of the photograph;
the Pb strip is the horizontal white region. }
 \label{figone}
 \end{figure}

The experiments were performed at UC San Diego (UCSD) and UC Berkeley
(UCB) on single crystals of $\rm Y Ba_2^{} Cu_3^{} O_{7-\delta}^{}$
grown at UCSD and the University of Illinois, respectively. Details of
the growth and characterization of sparsely twinned, single crystals
of $\rm Y Ba_2^{} Cu_3^{} O_{7-\delta}^{}$ with $c$-axis orientation
have been published previously \cite{maple,ginsberg}.  The large
dimensions of the domains, typically 200 x 200 $\mu{\rm m}^2$, allowed
us to position Josephson junctions across two twin domains separated
by a single twin boundary or even on a single domain. The UCSD
junction fabrication process is described in
Refs. \cite{dynesa,valles}.  Similar techniques were used at UCB
except in that an SiO insulating layer, rather than epoxy, was used to
define the junction geometry and a Pb$_{0.95}$In$_{0.05}$ rather than
Pb source was used to deposit the counterelectrode.  Figure 1(b) is a
photograph of junction B1 taken under a Nemarsky optical microscope; a
single YBCO twin boundary is clearly visible.  The dimension of the
junction perpendicular to the twin boundary is defined by the opening
in the SiO layer (L=$240\,\mu \rm m$).  The dimension of the junction
parallel to the twin boundary is determined by the width of the Pb
counterelectrode (W=$180\,\mu\rm m$).

We present results from eight samples, with parameters listed in Table
I.  To minimize trapped flux, the junctions were cooled slowly from
about $125\,$K to $4.2\,$K in a dewar surrounded by a $\mu$-metal
shield.  We measured $I_c({\bf B})$ with the magnetic field ${\bf B}$
applied in the plane of the junction at a variety of angles relative
to the twin boundary.  At UCSD, the samples were physically rotated
relative to the field.  At UCB, two perpendicular pairs of Helmholtz
coils were used to rotate the field relative to the sample, providing
angular control to $\pm 1^o$.  All of the junctions displayed low
leakage, superconductor-insulator-superconductor tunneling
characteristics similar to those seen in previous work
\cite{dynesa,dynesb,losueur,dynesd,kleiner}, including a
well-defined Pb energy gap ($\approx 1.4\, {\rm meV}$) and sharp Fiske
modes, indicating a high quality tunneling barrier.  All of the
junctions were in the small junction limit ($L/\lambda_J^{}<1$, where
$\lambda_J$ is the Josephson penetration depth), except for junctions
SD3 and B1 where $L/\lambda_J^{}\approx 1.2$ and $1.5$, respectively.
As shown in Table I, the $I_c^{\rm max}R_{N}$ products ($I_c^{\rm
max}$ is the maximum value of $I_c({\bf B})$, $R_{N}$ is the
resistance above the gap) lie between 0.5mV and 1mV, even though the
values of $I_c^{\rm max}$ and $R_{N}$ vary by more than an order of
magnitude.  This result suggests that variations in $I_c^{\rm max}$
and $R_{N}$ arise from variations in the properties of the tunneling
interface and that  the properties of the YBCO crystals do not change
substantially from sample to sample.  


Figures 2 and 3 contain the main result of this paper.  Figure 2 shows
$I_c({\bf B})$ for junction SD2 with the field parallel to the twin
boundary.  In addition to a deep local minimum in $I_c(B=0)$, we
observe a high degree of symmetry under reversal of both field and
current, indicating that the junction is free of significant trapped
flux.  Figure 3 shows $I_c({\bf B})$ for junction B2a as a function of
the angle $\phi$ between the twin boundary and $\bf B$. For
${\phi}=-90^{o}$, we observe a conventional, Fraunhofer-like pattern.
As we increase $\phi$, two local maxima in critical current develop on
either side of the $B=0$ axis, while the value of $I_{c}(B=0)$ remains
constant.  These two peaks have a maximum height at ${\phi}=0^{o}$, and
decrease again as we increase $\phi$ to +90$^{o}$.  The behavior shown
in Figs.\ 2 and 3 was observed in all junctions grown across a
twin boundary.

 \begin{figure}[b]
\epsfysize=7cm\epsfxsize=8cm\centerline{\epsfbox{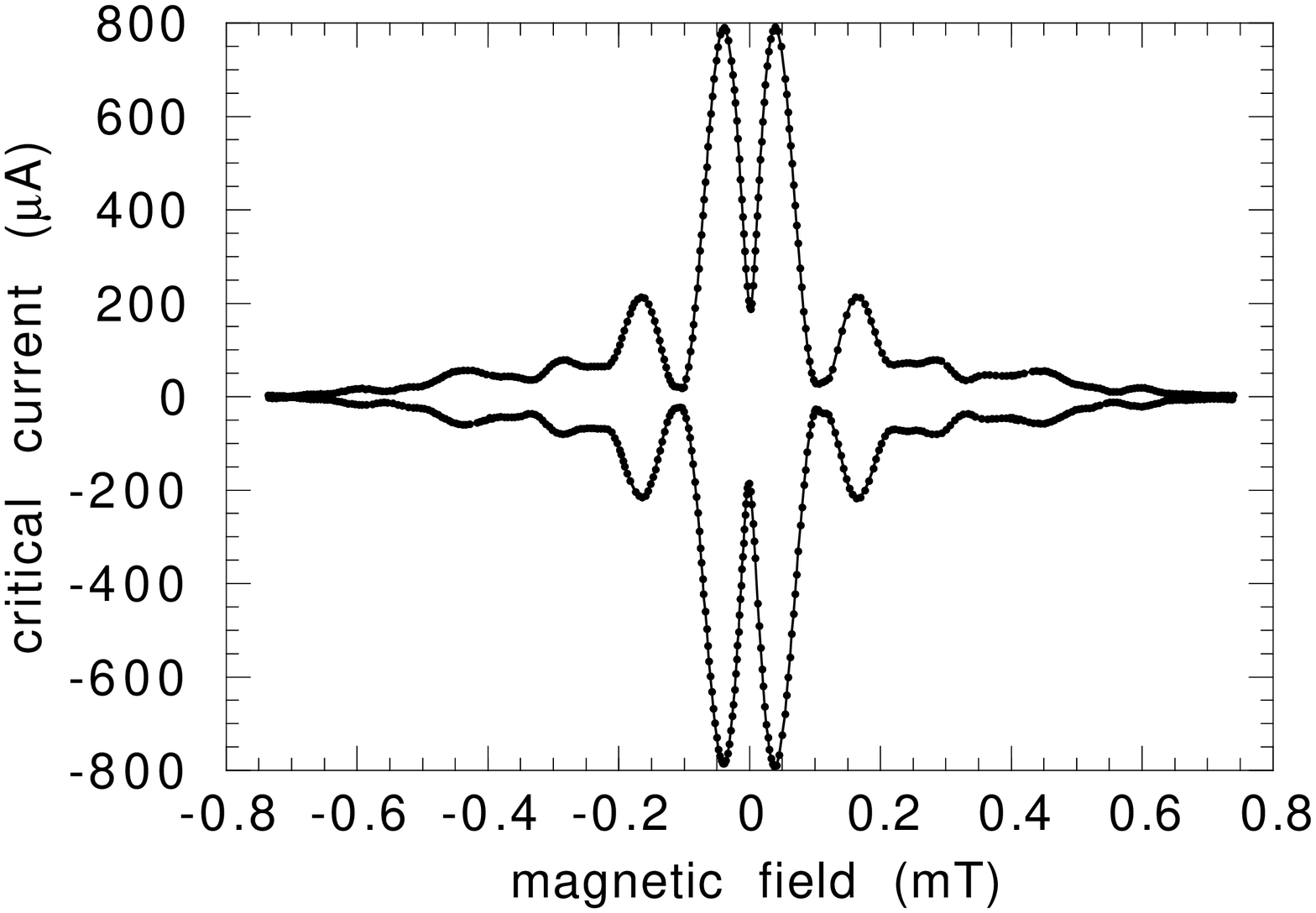}}
 \caption{
$I_c({\bf B})$ for junction SD3 ($\gamma=0.48$).  The magnetic field
is applied parallel to the twin boundary.}
 \label{figtwo}
 \end{figure}

The observed, reproducible dependence of the critical current on the
angle $\phi$ strongly suggests an interpretation in terms of a sign
reversal of the $s$-wave component of the YBCO order parameter across
the twin boundary.  We define the geometric asymmetry of the position
of the twin boundary within the junction as $\gamma \equiv A_1/A$,
where $A_1$ is the area of the smaller twin domain and $A=A_1+A_2$ is
the total area of the junction.  
For an $s$-wave component which
changes sign across the twin boundary, and for $\bf
B$ applied parallel to the boundary, 
we can write the critical current of a
small ($L \ll \lambda_J^{}$), uniform junction as
\begin{eqnarray}
I_c^2(\Phi,\gamma)&=&
\left({I_0\,\Phi_0/\pi\Phi}\right)^2
\;\Big[1+\cos^2\left({\pi\Phi/\Phi_0}\right)
 \\
& &
-\cos\left({2\gamma\pi\Phi/\Phi_0}\right) 
-\cos\left[{2(\gamma-1)\pi\Phi/\Phi_0}\right]\Big].
\nonumber
\end{eqnarray}
Here, $I_0$ is the maximum critical current of the junction with no
sign reversal, $\Phi$ is the magnetic flux penetrating the junction,
and $\Phi_0=hc/2e$ is the flux quantum.  For $\gamma=0$ or 1, Eq. (1)
yields the conventional result $I_c=I_0 |\sin(\pi\Phi/\Phi_0) /
(\pi\Phi/\Phi_0)|$.  For $\gamma=0.5$, we find $I_c=I_0
\sin^2(\pi\Phi/2\Phi_0) / |\pi\Phi/2\Phi_0|$, which is identical to
the result for the $d$-wave corner junction \cite{reviews,wollmanfm}.
We also calculated $I_c({\bf B})$ as a function of $\phi$; the
results for $\gamma=0.4$ are shown inset in Fig.\ 3.  The predicted
behavior is similar to our experimental observations: a continuous
variation as a function of $\phi$ from a local minimum at $B=0$ for
fields parallel to the twin boundary to a global maximum at $B=0$ for fields
perpendicular to the twin boundary.

 \begin{figure*}
\epsfysize=10cm\epsfxsize=12cm\centerline{\epsfbox{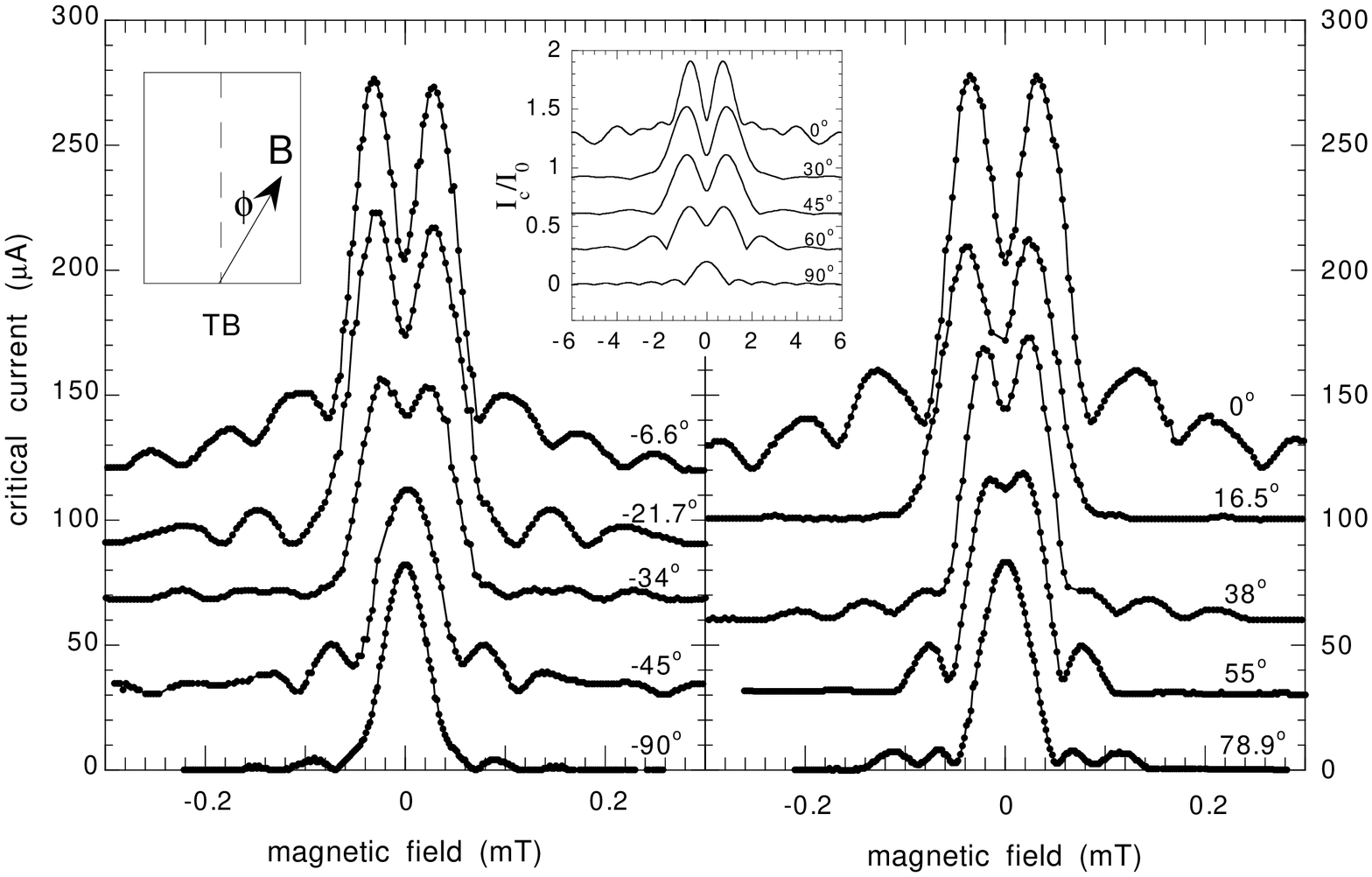}}
 \caption{
$I_c({\bf B})$ for junction B2a ($\gamma$=0.4) for 10 values of angle
$\phi$ relative to the twin boundary.  Successive plots are offset
vertically by 30 $\mu$A.  Center inset: computed values of $I_c/I_0$
{\it vs.}\ $\Phi/\Phi_0$ for a 
junction with a sign-reversal of the s-wave component and
$\gamma=0.4$.  Successive plots are
offset vertically by 0.3.}
 \label{figthree}
 \end{figure*}

Figure\ 4 shows $I_c({\bf B})$ for ${\phi}=0^{o}$ and 90$^{o}$ for
junction B2b grown on a {\it single} twin domain of the same crystal
as in Fig.\ 3.  We observed conventional, Fraunhofer-like behavior,
with a global maximum at $B=0$, for all values of $\phi$.  Because
YBCO is orthorhombic, the order parameter is expected to be of the
form $d+\epsilon s$ within a single domain.  The Pb $c$-axis junction
couples only to the $s$-wave component, so that a Fraunhofer-like
$I_c({\bf B})$ for all angles $\phi$, as seen in Fig.\ 4, is expected.
The contrast between the data shown in Fig.\ 4 and the behavior
observed in the other seven junctions makes it clear that the minima at
$B=0$ in those junctions are due to the presence of the twin boundary.

In Table I, we summarize $I_c({\bf B})$ for all eight junctions in
terms of the ratio $\eta\equiv I_c^0/I_c^{\rm max}$, where $I_c^0$ is
the critical current at zero field and $I_c^{\rm max}$ is the largest
critical current in a field applied parallel to the twin boundary.  We
also list $\eta_{th}$, the value of $\eta$ computed from Eq. (1) for
the measured values of $\gamma$.  For a conventional $s$-wave
junction, or a twin boundary junction in perpendicular applied fields,
$\eta_{th}=1$.  For an ideal, symmetric twin boundary junction
($\gamma=0.5$) in parallel applied fields, $\eta_{th}=0$.  For the
junction B1, with $\gamma=0.33$, the predicted and measured values of
$\eta$ are in good agreement.  On the other hand, for the one junction
(SD4) for which $\eta$ was measured to be zero, $\gamma$ was 0.42 and
$\eta_{th}=0.22$.  For the remaining five junctions with $0.40\le
\gamma \le 0.50$, the measured values of $\eta$ are consistently
higher than the predicted values.  The difference might be due to
self-field effects ({\it i.e.}, the junctions might not be small
enough compared to $\lambda_J^{}$ so that screening can be neglected).
However, Kirtley {\it et al.}\ \cite{kms} have shown that for
$L/\lambda_J^{}$ as large as 1 the expected $\eta_{th}$ for a
symmetric junction $(\gamma=0.5)$ is no more than $0.12$; thus
self-field effects appear to be too small to explain the
discrepancies.

Two possible sources of the discrepancy in $\eta$, which would also
explain the observation of non-zero supercurrents in heavily twinned
samples \cite{dynesa,dynesb,losueur,dynesd,kleiner}, are: (1) an
additional $s$-wave component to the order parameter induced by the
presence of the surface \cite{bahcall}, and (2) a localized
time-reversal breaking ($d+is$) state at the twin boundary, due to a
continuous rotation of the relative phase between $d$ and $s$
\cite{sigrist}.  We note that the data in Fig.\ 3 display some small
asymmetry in ${\bf B}\to -{\bf B}$ at intermediate angles.  However,
the asymmetry vanishes at $\phi=0^o$, to within the limit of our
resolution, whereas a time-reversal breaking state at the twin
boundary would predict the asymmetry to be a maximum at $\phi=0^o$.
We can use the symmetry of the data to place a limit on the degree of
time-reversal breaking at a single boundary: we find that the length
over which the phase angle rotates must be $\alt 4\, \mu{\rm m}$
\cite{konstbeth}.  Finally, we note that a non-uniform junction could
cause a discrepancy between the predicted and observed values of
$\eta$.   A spatially varying barrier height would mean that the
$\gamma$ inferred from the relative areas of the domains might not
reflect the actual asymmetry in critical current magnitudes.

Could the large observed minima in $I_c$ at $B=0$ arise from trapped
vortices in a purely $s$-wave superconductor?  A vortex containing
flux $\Phi_0$ trapped within the plane of the twin boundary, at an
angle $0^o<\theta<90^o$ relative to the $c$-axis, would have the same
effect as a phase shift of $2\pi\sin\theta$ at the boundary.  We have
simulated the effect of such a vortex: the expected $I_c({\bf B})$
pattern is asymmetric in ${\bf B}\to -{\bf B}$ except when the
component of the flux perpendicular to the plane of the junction is
exactly $\Phi_0/2$, that is, when $\theta=30^o$.  A trapped vortex out
of the plane of the twin boundary exhibits the same behavior, except
that the minima in $I_c({\bf B})$ occur at magnetic field angles parallel
to the direction of the vortex.  The highly reproducible symmetry of
the data, from sample-to-sample and from run-to-run on a given sample,
indicates that trapped magnetic flux is an extremely unlikely
explanation of the data.

 \begin{figure}[b]
\epsfysize=7cm\epsfxsize=8cm\centerline{\epsfbox{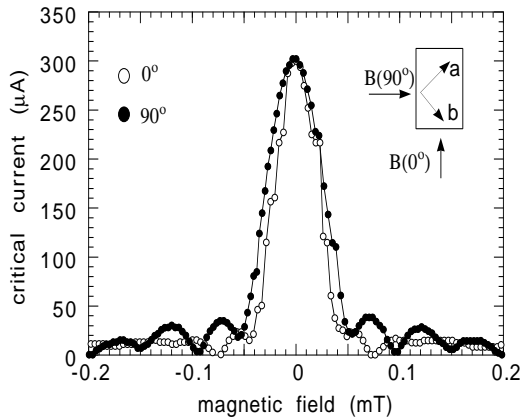}}
 \caption{
$I_c({\bf B})$ for junction B2b, grown on a single twin domain,
showing conventional, Fraunhofer-like behavior for two orthogonal
field directions.}
 \label{figfour}
 \end{figure}

In conclusion, we find that junctions grown across a single twin
boundary of YBCO crystals consistently exhibit a local minimum in
$I_c({\bf B})$ at zero magnetic field for fields applied in the plane
of the crystal parallel to the twin boundary.  As the magnetic field
is progressively rotated in the plane of the junction, the value of
$I_c(0)$ remains constant while the maxima on either side diminish.
The smooth variation of $I_c({\bf B})$ with magnetic field direction,
its symmetry with respect to the reversal of magnetic field and bias
current, and the reproducibility of the results imply that trapped
flux is an extremely unlikely source of the minima.  The results are
entirely consistent with a predominantly $d_{x^2-y^2}$ pairing
symmetry with a sign reversal of the $s$-wave component across the
twin boundary.

We thank S.\ Kivelson, R.\ Klemm, K.\ A.\ Moler, D.\ J.\ Scalapino,
S.\ I.\ Woods, and Peng Xiong for beneficial discussions during the
course of this experiment.  The work at UCSD was supported by AFOSR
Grant No.\ F4962-092-J0070, NSF Grant No.\ DMR 91-13631, and DOE Grant
No.\ DE-FG03-86ER-45230; at UCB by the Director, Office of Energy
Research, Office of Basic Energy Sciences, Materials Science Division
under DOE Grant No.\ DE-AC03-76SF-00098; and at UIUC by NSF Grant No.\
DMR 91-20000 through the Science and Techology Center for
Superconductivity.


\begin{table}
\caption{Junction length L and width W, asymmetry parameter ${\gamma}$,
zero-field
critical current $I_c^0$, maximum critical current 
in a parallel field ($\phi=0^o$) $I_c^{\rm max}$, 
$\eta_{th}\equiv I_c^0/I_c^{\rm max}$ calculated from Eq. (1), 
$\eta$ from experiments, 
and $I_c^{\rm max} R_N^{}$ product for
eight $c$-axis YBCO/Pb junctions.}
\begin{tabular}{cccdddccd}
 & $L$ & $W$ & $\gamma$ & $I_c^0$ & 
 $I_c^{\rm max}$ &
 $\eta_{th}$ & $\eta$ &  
 $I_c^{\rm max} R_N^{}$ \\ 
 & ($\mu$m) & ($\mu$m)
 & & ($\mu$A) &($\mu$A) &  & & (mV)\\
\tableline
SD1 & 250 & 445 & 0.45 & 226 & 396 & 0.14 & 0.57 & 1.03 \\ 
SD2 & 150 & 330 & 0.47 & 193 & 815 & 0.08 & 0.24 & 0.90 \\ 
SD3 & 250 & 280 & 0.48 & 300 & 2010 & 0.06 & 0.15 & 1.03 \\ 
SD4 & 250 & 350 & 0.42 & 0 & 28 & 0.22 & 0 & 0.90\\ 
B1 & 240 & 180 & 0.33 & 780 & 1550 & 0.49 & 0.50 & 0.68\\ 
B2a & 175 & 240 & 0.40 & 82 & 158 & 0.28 & 0.52 & 0.51\\ 
B2b & 180 & 250 & 0.0 & 300 & 300 & 1.00 & 1.00 & 0.75 \\ 
B3 & 160 & 250 & 0.50 & 10 & 37.6 & 0.0 & 0.27 &  1.00 \\ 
\end{tabular}
\end{table}

\bigskip


\begin{references}
\bibitem{signch} 
Across a twin boundary, the $a$- and $b$-axes rotate by $90^o$ so that
the relative sign between $s$- and $d$-components must change.  The
``dominant'' component is the one whose phase does not change across
the boundary.  That component is preserved throughout the sample in the
heavily twinned limit.
\end{references}
\end{document}